\documentclass{article}
\usepackage[utf8]{inputenc}
\usepackage{authblk}
\usepackage{setspace}
\usepackage{float}
\usepackage[margin=1.25in]{geometry}
\usepackage{graphicx}
\usepackage{physics}
\graphicspath{ {./figures/} }
\usepackage{subcaption}
\usepackage{amsmath}
\usepackage[backend=biber,sorting=ynt]{biblatex}
\usepackage{xcolor}
\title{Bayesian optimisation approach to quantify the effect of input parameter uncertainty on predictions of numerical physics simulations}
\author[1]{Samuel G. McCallum}
\author[1]{James E. Lerpini\'{e}re}
\author[2] {Kjeld O Jensen}
\author[3,4*] {Pascal Friederich} 
\author[1*]{Alison B Walker}
\affil[1] {Department of Physics, University of Bath, UK}
\affil[2]{Applied Research, British Telecommunications, UK}
\affil[3]{Institute of Theoretical Informatics, Karlsruhe Institute of Technology, Germany}
\affil[4]{Institute of Nanotechnology, Karlsruhe Institute of Technology, Germany}
\affil[*]{Corresponding authors. Email: pascal.friederich@kit.edu, pysabw@bath.ac.uk}

\date{}
\begin{document}
\maketitle
\begin{abstract}
An understanding of how input parameter uncertainty in the numerical simulation of physical models leads to simulation output uncertainty is a challenging task.  Common methods for quantifying output uncertainty, such as performing a grid or random search over the model input space, are computationally intractable for a large number of input parameters, represented by a high-dimensional input space. It is therefore generally unclear as to whether a numerical simulation can reproduce a particular outcome (e.g. a set of experimental results) with a plausible set of model input parameters. Here, we present a method for efficiently searching the input space using Bayesian Optimisation to minimise the difference between the simulation output and a set of experimental results. Our method allows explicit evaluation of the probability that the simulation can reproduce the measured experimental results in the region of input space defined by the uncertainty in each input parameter. We apply this method to the simulation of charge-carrier dynamics in the perovskite semiconductor methyl-ammonium lead iodide MAPbI$_3$ that has attracted attention as a light harvesting material in solar cells. From our analysis we conclude that the formation of large polarons, quasiparticles created by the coupling of excess electrons or holes with ionic vibrations, cannot explain the experimentally observed temperature dependence of electron mobility.
\end{abstract}
The following article has been submitted to APL Machine Learning. After it is published, it will be found at
https://publishing.aip.org/resources/librarians/products/journals/

\section{Introduction}
Numerical simulation of physical models typically require a large number of input parameters that must be derived from previous theoretical calculations or measured experimentally \cite{Car85, Payne92}. It is therefore inevitable that model input parameters suffer from uncertainty, and the effect of this uncertainty on the simulation output is not often discussed. Quantifying the functional relationship between the model input parameters and simulation output, in the $N$-dimensional input space where a set of $N$ model input parameters is a point in this space, is computationally challenging due to the combination of a high-dimensional input space and expensive numerical simulation times \cite{Caron19}. The effect of input uncertainty is thus routinely ignored given that naive methods for exploring the input space, such as grid or random search, require a computationally intractable number of simulations. It is therefore difficult to establish whether it is possible to generate a particular outcome, e.g. matching a set of experimental results, from a numerical simulation with a plausible set of input parameter values.

Machine learning, specifically Bayesian Optimisation \cite{Snoek12, Garnett22}, can be used to efficiently search the input space in the region defined by the uncertainty in each input parameter, while minimising the difference between the simulation output and the experimentally measured result. During optimisation a probabilistic function is learned that maps the input parameters to the simulation output. This function explicitly evaluates the probability that the simulation can reproduce the experimental result in the region of input parameter space defined by the uncertainty. With this approach, informed conclusions can be drawn from the simulation output that allow for the uncertainty present in the input parameters. The method presented here falls into the wider field of statistical emulation for sensitivity and uncertainty analysis \cite{Oakley02, Oakley04, OHagen06}.

In the field of semiconductor physics, previous methods have attempted to manually minimise the error between simulation and experiment in the context of device parameter estimation for organic light-emitting diodes (OLEDs) \cite{Neukom19, Jenatsh18, Jenatsh20}. However, a machine-learning method for automatically quantifying the effect of input parameter uncertainty on simulation prediction by minimising the error between simulation and experiment has not been established. A related problem, of utilising simulation prediction to find device parameters from experimental measurements, was demonstrated by Knapp et al. \cite{Knapp21}.

We demonstrate the method with application to the simulation of charge-carrier dynamics in lead-halide perovskites, LHPs. LHPs as light-harvesting layers in solar cells have been the focus of intense research activity due to power conversion efficiency increases of 9-27$\%$ since 2009 \cite{Jena19, Green14} and low-energy low-cost manufacturing \cite{Razza16}. However, a comprehensive theoretical understanding of the mechanisms responsible for a number of their electronic properties is currently lacking \cite{Egger18, Herz18}. Specifically, the fundamental physical mechanisms controlling transport of photogenerated charge carriers is actively debated, see for example the measurements of Kobbekaduwa et al \cite{Kobbekaduwa23} on ultrafast dynamics that address the carrier mobility dependence on temperature in LHPs. Mesoscale simulation models of charge transport in LHPs, such as BoltMC - an ensemble Monte Carlo approach using Boltzmann transport theory \cite{Irvine21}, can provide insight into these mobility-limiting mechanisms.

Here we investigate the possibility that the formation of large polarons \cite{Frost17}, quasi-particles resulting from the coupling of charge carriers with the surrounding polarised ionic lattice, affects the temperature dependence of the carrier mobility in methyl-ammonium lead iodide, MAPbI$_3$.  A large polaron is an itinerant particle, with a mobility that decreases with temperature \cite{Irvine21}. Prior theoretical work has utilised many approaches for calculating this dependence \cite{Frost17, Schlipf18, Okamoto74}, parameterised from electronic structure calculations and/or experimental measurements of material parameters. Yet, the effect of uncertainty in these material input parameters has been largely ignored. Here we show, with calculations from BoltMC, that the temperature dependence of polaron mobility is a non-linear function of the input parameters in the region of input space defined by the uncertainty in each parameter. This finding indicates that strong conclusions derived from a single point (i.e. a single set of material input parameters) in the input space are unjustified. Bayesian Optimisation has thus been used to efficiently explore the effect of uncertain inputs on simulation predictions of the temperature dependence of polaron mobilities in comparison to experiment. This approach strengthens any conclusions drawn from the simulation output as the effect of input parameter uncertainty has been considered. And, in context, aids our understanding of the fundamental physical mechanisms influencing charge transport in MAPbI$_3$.

\section{Results and Discussion}
\subsection{Polaron mobility temperature exponent}
\label{subsec:resultspolmob}
\begin{figure*}[!ht]
\includegraphics[scale=1.7]{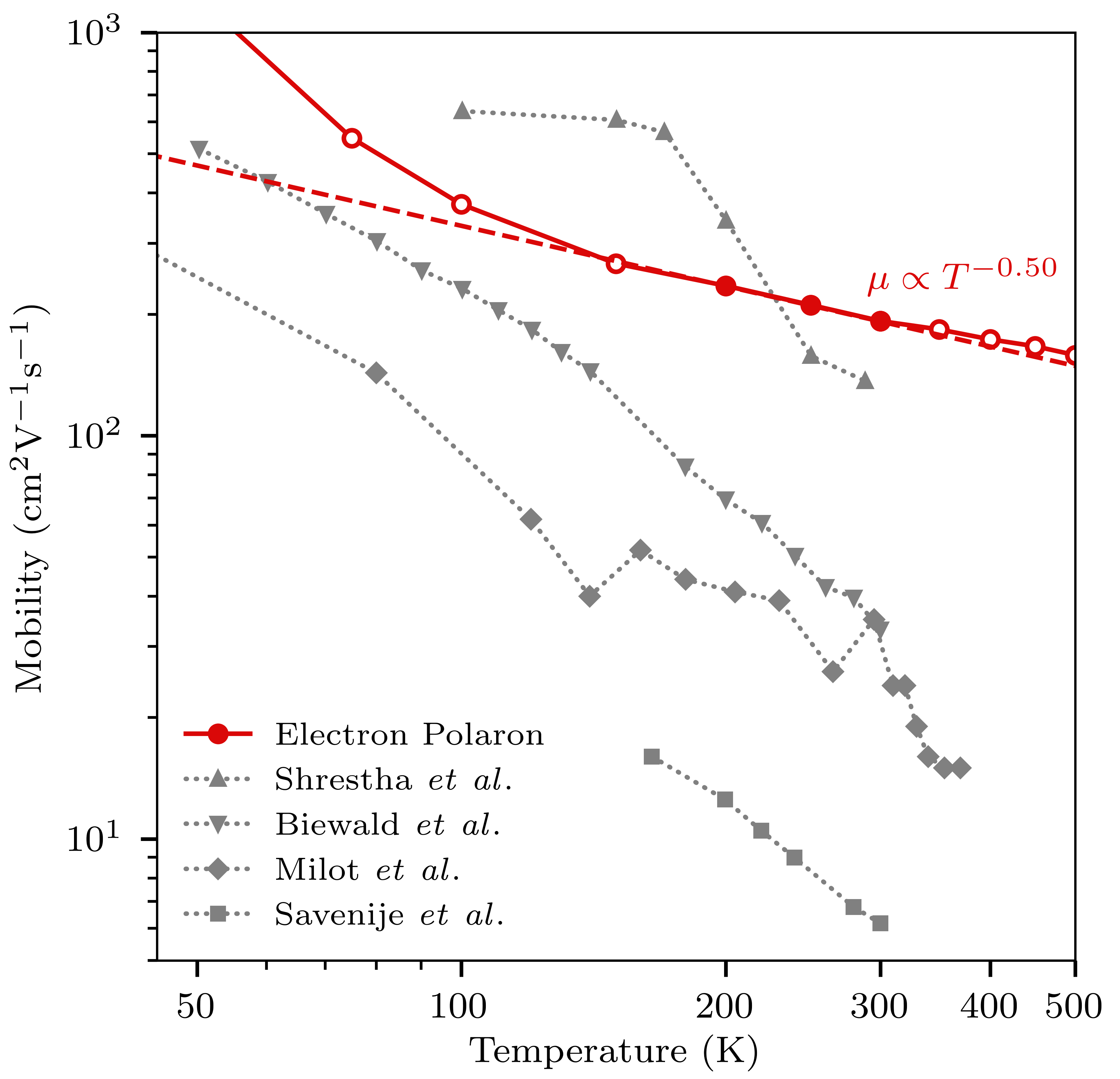}
\caption{Temperature dependent mobilities of electron polarons in MAPbI$_3$ associated with material parameters found to minimise the polaron mobility temperature exponent (see Table 1). Solid red points show mobilities used during optimisation; open red points show mobilities calculated after optimisation to confirm exponential temperature dependence (for $T \geq 200K$). Errors in the mobility ensemble average are comparable to the data point size. Experimental data points are shown in grey with the symbols shown in the legend and are linked by dashed lines: Shrestha \cite{Shrestha18}, Biewald \cite{Biewald19}, Milot \cite{Milot15}, Savenije \cite{Savenije14}.}
\label{fig:mobilitiesvst}
\end{figure*}

Assuming the polaron mobility varies with temperature as a power law, Bayesian Optimisation was used to minimise the absolute value of the difference between the predicted temperature exponent and an experimentally measured exponent of $-1.5$  (see Figure \ref{fig:mobilitiesvst}). This power law dependence was assumed to be linear in the high-temperature limit, found in previous studies to be for temperatures $>200$K \cite{Irvine21}. The predicted temperature exponent was determined from the polaron mobility at three temperatures towards the lower temperature end of this limit ($200$K, $250$K and $300$K) in order for the procedure to be computationally efficient; the simulation time increases with temperature due to increased scattering at higher temperatures.

Using the approach described in section \ref{sec:methods}, 1269 simulations were required from 423 points (including initial set) in the 6-D simulation input parameter space at temperatures $200$K, $250$K and $300$K. The minimum temperature exponent of polaron mobility in MAPbI$_3$ was found to be {$-0.50 \pm 0.04$}. This prediction resulted from optimisation in the hypervolume of input space defined by the uncertainty in each of the six material simulation input parameters. The optimisation procedure was terminated due to Gaussian Process Regression (GPR) predicting the probability of obtaining a temperature exponent equal to or more negative than $-1.5$ of $\sim 10^{-24}$. Table 1 shows the material parameters associated with the minimum temperature exponent found during minimisation. All material parameters lie within the assumed uniform uncertainty, except the effective mass $m^{*}$ which was found to minimise the temperature exponent at 0.12 $m_e$ (bottom of uncertainty range). This suggests that the temperature exponent may be further minimised by decreasing $m^{*}$. On increasing the prior uncertainty range to $\pm 40\%$ and allowing the minimisation procedure to continue, it was found that the temperature exponent may be further minimised to $-0.54$ with $m^{*} = 0.098 m_e$. This value of $m^{*}$ is well within the $\pm 40\%$ range, indicating that the temperature exponent may not be further minimised by decreasing $m^{*}$ (within this range). Additionally, a temperature exponent of $-0.54$ is within one standard error of the temperature exponent found within the $\pm 20\%$ uncertainty range and therefore does not affect any conclusions drawn below. Finally, increasing the material parameter uncertainty range to $\pm 40\%$ also leads to material parameters that are un-physical for MAPbI$_3$ (see \cite{Frost17, Brivio13, Brivio14, Lee18}).

An accurate prediction of the probability of reproducing a temperature exponent of $-1.5$ relies strongly on the accuracy of the output variance predicted by GPR. A measure of this accuracy is easily obtained and is described as follows. In the 6-D simulation input space, $100$ points were randomly generated and the temperature exponent of the polaron mobility calculated. Using the trained Gaussian Process Regressor, the predicted normal distribution (see Equation 3) over the temperature exponent for the $100$ points in input space was determined and $\pm 2 \sigma, \pm 3 \sigma$ $(95.5\%, 99.7\%)$ confidence bounds were calculated. $88\%$ of simulated temperature exponents, for the $100$ points in input space, fell within $95.5\%$ predicted bounds, with $100\%$ of simulated temperature exponents within $99.7\%$ bounds. This indicates that the uncertainty of the GP (Gaussian Process) model is reasonably accurate and does not require further uncertainty calibration \cite{Kumar19}.

The accuracy of predicted GP model uncertainties provides strong evidence that the probability of obtaining a temperature exponent more negative than $-1.5$ is indeed vanishingly small, $\sim 10^{-24}$. Therefore, this modelling indicates that polaron formation in MAPbI$_3$ cannot explain the observed temperature dependence of electron mobility, even once the effect of a large uncertainty in the material simulation input parameters is considered. Other mechanisms that regulate charge-carrier dynamics in MAPbI$_3$ must be investigated, such as trapping and recombination, as polaron formation has been shown to be more limited in its effect on electron mobilities than previously hypothesised in the literature \cite{Zhu15, Mante17, Zhang17}.

\begin{table}[H]
\caption{\label{arttype}Material parameters for electrons in MAPbI$_3$. Mean values sourced from \emph{ab initio} calculations reported in the literature and summarised in reference \cite{Irvine21}. Also shown are parameter values associated with the minimum discrepancy found during optimisation between the temperature exponent calculated from simulation and an experimental exponent of $-1.5$. Here m$_e$ is electron mass, $\varepsilon_0$ vacuum electric permittivity.}
\footnotesize\rm
\begin{tabular*}{\textwidth}{@{}l*{15}{@{\extracolsep{0pt plus12pt}}l}}
\hline
Parameter & Mean value & Minimum exponent discrepancy value \\
\hline
Conduction band effective mass $m^{*}$ & 0.15 $m_e$ & 0.12 $m_e$ \\
Polar optical phonon frequency $\omega_{\mathrm{pop}} / 2 \pi$ & 2.25 THz & 2.67 THz\\
Low frequency permittivity $\varepsilon_{\mathrm{LF}}$ & 25.7 $\varepsilon_{0}$ & 27.1 $\varepsilon_{0}$\\
High frequency permittivity $\varepsilon_{\mathrm{HF}}$ & 4.5 $\varepsilon_{0}$ & 5.4 $\varepsilon_{0}$\\
Acoustic deformation potential $\Xi$ & 2.13 eV & 1.7 eV\\
Elastic constant $c_{\mathrm{L}}$ & 32 GPa & 32 GPa\\
\hline
\end{tabular*}
\end{table}

\subsection{Band electron and polaron mobility at 298K}
Additionally, GPR was fit to both the band electron and polaron mobility as a function of the 6-D simulation input space at 298K, with mobilities calculated as described in section \ref{subsec:simpoldyn}. A training set of 1000 simulations was obtained from randomly generated sets of simulation input parameters and their calculated band electron and polaron mobilities. A mean absolute percentage error of $0.5\%$ was achieved for both model predictions of band electron and polaron mobility on an unseen test set of 200 simulations. The trained length-scales of the covariance function of the GP for the two models provide information regarding the physical scattering mechanisms present in MAPbI$_3$. Specifically, a measure of the sensitivity of the band electron and polaron mobility to each of the six input parameters (normalised by removing mean and scaling to unit variance) is quantified by the inverse of the length-scale for each parameter (see Figure 2) \cite{Chaudhuri19}.

\begin{figure*}[h!]
\includegraphics[scale=1.45]{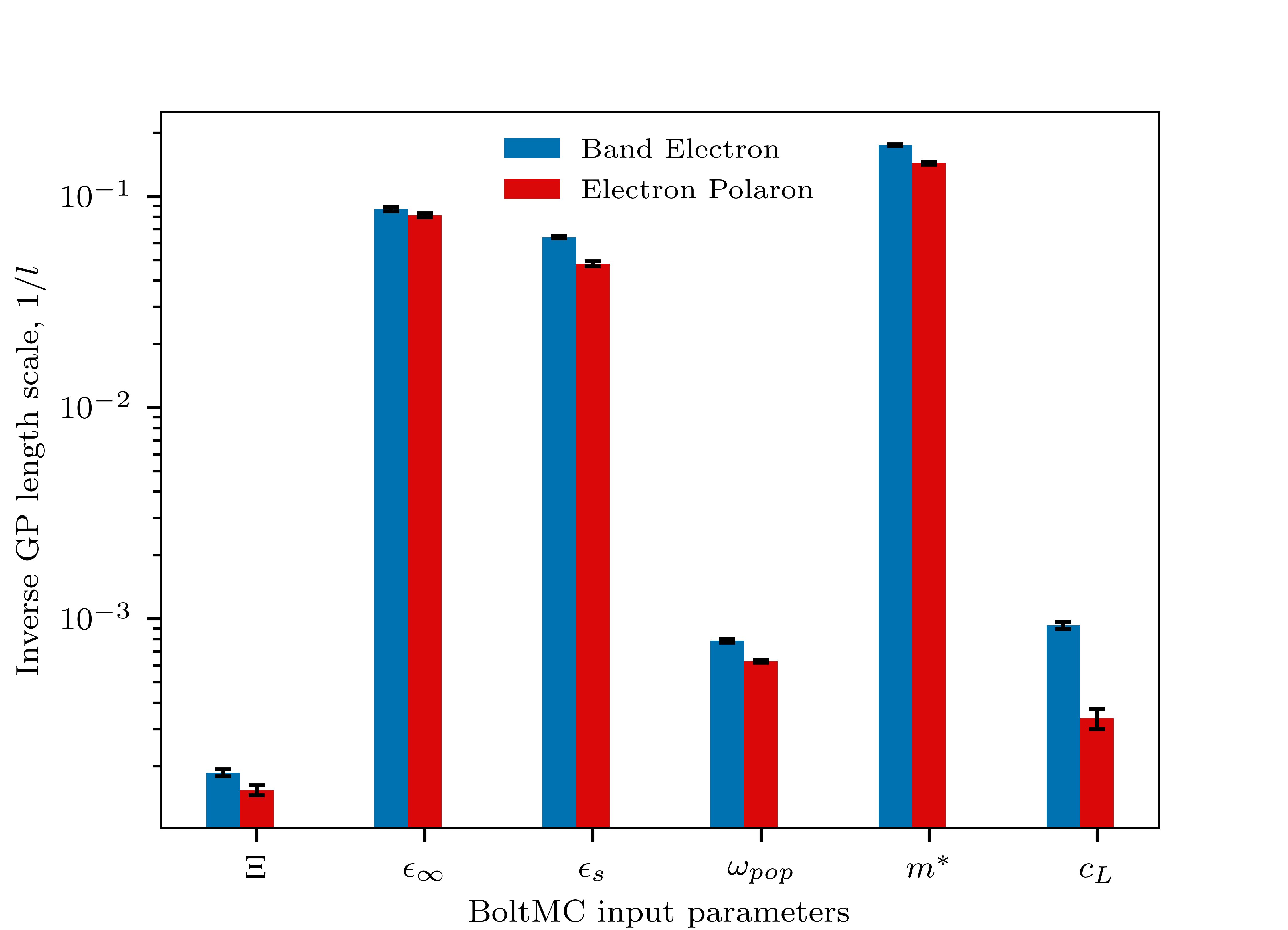}
\caption{The inverse length-scales of covariance function (RBF) for each simulation input parameter, defined by mean values in Table 1, taken from two trained GPs predicting band electron and polaron mobility from simulation input parameters at 298K. The inverse length-scales are plotted such that larger values indicate larger output (mobility) sensitivity to that parameter. Error bars show standard error in the mean in black and were calculated by 10-fold cross validation over the training set.}
\end{figure*}
An understanding of the effect of a small perturbation to an input parameter on the band electron and polaron mobility is therefore obtained, here providing insight into the physical mechanisms controlling polaron mobility in lead-halide perovskites. For MAPbI$_3$, this analysis suggests that the polaron mobility is less sensitive to a perturbation in any input parameter than seen for bare band electron mobility. Given the six input parameters define the polaron scattering rates for acoustic phonons, optical phonons and ionised impurities \cite{Irvine21}, this in turn suggests that the polaron mobility is less sensitive to perturbations in the scattering rates.

Furthermore, comparison across the magnitude of length-scales for each parameter (note that comparison between length-scales for different parameters is possible only as the input data was normalised before fitting GPR) indicate which perturbed scattering rates the electron mobility is most sensitive to, within the uncertainty in each input parameter.  For example, a perturbation in the conduction band effective mass $m^*$ has the largest influence on the mobility. This can be rationalised from simple Drude theory $\mu = \tau(m^*) / m^*$, where the mobility is mediated by the magnitude of $m^*$ directly as well as implicitly through the relaxation time $\tau(m^*)$, arising from the dependence of the scattering rates on $m^*$. The effective mass can be seen to strongly regulate electron mobility, a result that is confirmed by this analysis.

Large mobility sensitivity to both low and high frequency permittivity, $\epsilon_s$ and $\epsilon_\infty$, is indicated by large inverse length-scales for both parameters. This suggests that the band electron/polaron mobility in MAPbI$_3$ has increased sensitivity to a perturbation in scattering rates strongly dependent on these parameters; namely, the scattering rates for polar optical phonons and ionised impurities due to their functional dependence on $(1/\epsilon_\infty - 1/\epsilon_s)$ and $1/\epsilon_s^2$, respectively \cite{Jacoboni89}. Walsh \cite{walsh22} discussed the importance of dielectric screening for defect tolerance of perovskites, and, we infer, to charge transport. While, despite the dependence of the acoustic phonon scattering rate on $\Xi^2$ and $1/c_L$, the mobility sensitivity (inverse length-scale) to each parameter ($\Xi$, $c_L$) remains low - suggesting that a perturbation in the acoustic phonon scattering rate has a reduced effect on the band electron mobility. This effect on mobility is significantly decreased for polarons, indicated by the smaller inverse length-scale. Indeed, previous studies of polaron formation in MAPbI$_3$ have concluded that scattering due to acoustic phonons is the mechanism most significantly decreased in comparison to bare band electrons \cite{Irvine21}, a result confirmed by this analysis. 

However, one must be careful in the physical insight derived from an analysis of the input parameter length-scales. Specifically, it is not clear that an input parameter with a larger length-scale has reduced significance in determining the band electron/polaron mobility; only that a perturbation in the parameter, and therefore perturbation in the scattering rate, has a smaller effect on the mobility. For example, it may be found that the magnitude of electron mobility is not significantly changed with a perturbation in one scattering rate. Yet this does \underline{not} mean that electron mobility would be unchanged if this scattering mechanism was not present in the simulation, only that its contribution to the mobility is approximately constant over the range that the material parameters were varied.

Finally, this analysis is useful for guiding future uncertainty reduction, whereby reducing uncertainty in input parameters with smaller length scales will result in a greater decrease in the uncertainty over the simulation output. Similarly, an analysis of the length-scale for each input parameter may be used to determine a reduction in dimensionality of the simulation input space, by excluding parameters with comparatively large length-scales, in turn improving the efficiency of search/optimisation methods.

\section{Methods}
\label{sec:methods}
\subsection{Efficient simulation input space search}
\label{subsec:efficient}
Global optimisation of an expensive-to-evaluate 'black box' function is a common problem across many disciplines \cite{Shan10}. Here, the 'black box' function refers to a numerical simulation of a parameterized physical model. Bayesian Optimisation \cite{Snoek12, Garnett22} is frequently used in this setting and utilises a learned probabilistic model of the function that is being optimised. No knowledge of the functional form is assumed - e.g. the function is not assumed to be convex and gradients are not directly accessible. GPR has been used to model the functional relationship between the simulation input parameters $\mathbf{x}$ and output $g(\mathbf{x})$. GPs \cite{Rasmussen06} can be viewed as specifying a distribution over functions,
\begin{equation}
    g(\mathbf{x})\sim\mathcal{GP}(m(\mathbf{x}),k(\mathbf{x},\mathbf{x'})),
\end{equation}
 where $m(\mathbf{x})$ defines the mean function and $k(\mathbf{x},\mathbf{x'})$ the covariance function of the GP. The radial basis function (RBF) was used in this work to calculate $k(\mathbf{x},\mathbf{x'})$, specifying the $i$'th row and $j$'th column of the covariance matrix with the addition of additive noise along the diagonal,
\begin{equation}
    k(\mathbf{x}_i,\mathbf{x}_j)=\sigma_f^2\exp\Big(-\frac{|\mathbf{x}_i-\mathbf{x}_j|^2}{2l^2}\Big) + \sigma_n^2 \delta_{ij}.
\end{equation}
The free parameters of the RBF, $\sigma_f^2$ and $l$, define the output signal variance and input length-scales respectively and $\sigma_n^2$ specifies the variance of additional Gaussian measurement noise. These parameters are optimised during training to maximise the log-marginal likelihood of the data. he choice of covariance function quantifies certain modelling assumptions that the underlying numerical function is expected to obey, such as smoothness, non-stationarity and periodicity \cite{Gelman13}. The RBF kernel encodes a prior assumption that the output of the function to be approximated smoothly varies with respect to function input parameters.

For a GP, the predicted marginal distribution of any single function value $y$ (simulation output) is univariate normal \cite{Rasmussen06}
\begin{equation}
    p(y|\mathbf{x}) = \mathcal{N}(y; \mu, \sigma^2), \hspace{5mm} \mu = m(\mathbf{x}), \hspace{5mm} \sigma^2=k(\mathbf{x},\mathbf{x}).
\end{equation}
Bayesian Optimisation utilises this predicted distribution by determining the optimal next input location with which to query the ground truth function (here, simulation model), such that the probability of finding an optimum is maximised. Specifically, an acquisition function $\alpha(\mathbf{x})$ is defined over the input space and the next input location with which to query the ground truth function is determined by maximising this acquisition function,
\begin{equation}
    \mathbf{x}_{n+1}=\arg \max_\mathbf{x}\alpha(\mathbf{x}).
\end{equation}
Here, the acquisition function is defined by the probability of improvement (PI), and if minimisation of the ground truth function is desired, is given by \cite{Rasmussen06}

\begin{equation}
    \alpha(\mathbf{x})= P(g(\mathbf{x}) \leq g(\mathbf{x}^+) - \kappa),
\end{equation}
where $\mathbf{x}^+$ represents the input location associated with the minimum function value observed thus far during optimisation and $\kappa$ is a hyper-parameter that defines the exploration-exploitation trade-off of the optimisation policy. 

In this application, the function minimised during optimisation is the absolute value of the difference between predicted simulation output and the experimental result. The optimisation procedure can be terminated once either a simulation output is found that reproduces the experimental result below a pre-specified error, or the probability of obtaining an output equal to the experimental result falls below a threshold. This probability can be explicitly evaluated from predicted variance of the simulation output from GPR at each point in the input space; a method for determining the accuracy of the predicted variances is reported in section 3. Note that the probability of reproducing the experimental result may increase or decrease on the addition of a further simulation. The threshold probability must therefore be defined conservatively (i.e. sufficiently low) to avoid incorrect early stopping and/or performing a minimum number of iterations before terminating due to low probability.

\subsection{Simulation of polaron dynamics}
\label{subsec:simpoldyn}
To demonstrate this method, polaron dynamics were simulated by solving an augmented form of Kadanoff's semiclassical Boltzmann transport equation \cite{Kadanoff63} using the ensemble Monte Carlo method \cite{Jacoboni89, Fawcett70}. For an ensemble of polarons subject to a constant electric field $\mathbf{E}$, the Boltzmann transport equation is given by
\begin{equation}
    \frac{\partial f}{\partial t} - \frac{e}{\hbar}\mathbf{E} \cdot \frac{\partial f}{\partial \mathbf{k}} = \Big( \frac{\partial f}{\partial t} \Big)_{pop}+\Big( \frac{\partial f}{\partial t} \Big)_{aco}+\Big( \frac{\partial f}{\partial t} \Big)_{imp},
\end{equation}
where $\mathbf{k}$ is the polaron wavevector and $f(\mathbf{k},t)$ defines the one-particle distribution function. The partial derivatives on the right hand side of Equation (6) represent the change to the distribution function due to the scattering of polarons by optical phonons (pop), acoustic phonons (aco), and ionised impurities (imp). The scattering rates for each of these three mechanisms were calculated using Fermi's golden rule,
\begin{equation}
    S(\mathbf{k}_i \rightarrow \mathbf{k}_j) = \frac{2\pi}{\hbar}|\bra{\mathbf{k}_f}H'\ket{\mathbf{k}_i}|^2\delta (\epsilon_f-\epsilon_i-\Delta\epsilon),
\end{equation}
where $H'$ is the time-dependent perturbing Hamiltonian for each scattering mechanism in the bulk of a polar semiconductor \cite{Jacoboni89}. $\epsilon_i$, $\epsilon_f$ are the energies of the initial and final states respectively and $\Delta \epsilon$ represents the quanta of energy exchanged as a result of the scattering. A derivation of the scattering rates for the three scattering mechanisms considered here for both band electrons and polarons can be found in \cite{Irvine21}.

Calculation of the scattering rates requires the polaron eigenstates $\ket{\mathbf{k}}$. These eigenstates were obtained from the Feynman model of a polaron where an electron is coupled to a second particle via a harmonic potential that represents the cloud of virtual phonons associated with the surrounding polarised ionic lattice \cite{Feynman55}. The Hamiltonian of this system is given by \cite{Kadanoff63}
\begin{equation}
    \begin{split}
    H_F &= \frac{\hbar^2 |\mathbf{k_e}|^2}{2m_e^*}+\frac{\hbar^2 |\mathbf{k}_c|^2}{2m_c^*} + \frac{1}{2}\kappa(\mathbf{r}-\mathbf{r}_c)^2 \\
    &= \frac{\hbar^2 |\mathbf{k}|^2}{2m^*} + \hbar\omega_{osc}\sum_{i=1}^3 \Big( (a_{osc})_i^\dag (a_{osc})_i + \frac{1}{2}\Big)
    \end{split}
\end{equation}
where $\mathbf{r}, \mathbf{k}$ and $m^*$ are the position, wavevector and effective mass for the electron (subscript 'e'), phonon cloud (subscript 'c') and polaron (no subscript). The polaron's internal harmonic oscillator state is defined by the angular frequency $\omega_{osc}$, with $(a_{osc})_i^\dag$ and $(a_{osc})_i$ representing the ladder operators for this harmonic oscillator system for the three cartesian directions, indexed by $i$.

Finally, steady-state solutions to the Boltzmann transport equation were obtained by the ensemble Monte Carlo method \cite{Jacoboni89, Fawcett70} (additional information of this calculation and its computational implementation can be found in \cite{Irvine21}) and the polaron mobility $\mu$ determined from the ensemble average of the polaron wavevector,
\begin{equation}
    \mu = \frac{\hbar |\mathbf{\overline{k}|}}{|\mathbf{E}|m^*}.
\end{equation}
Band electron dynamics were also simulated and the mobility calculated as described above but with the band electron state $\mathbf{k_e}$ determined from the effective mass Hamiltonian,
\begin{equation}
    H_0 = \frac{\hbar^2|\mathbf{k_e}|^2}{2m_e^*}.
\end{equation}

\subsection{Polaron dynamics under input uncertainty}
\label{subsec:polarondyn}

To calculate the polaron scattering rates, six material parameters were required that define the semiconductor to be simulated and can be viewed as specifying a 6-D input space. These parameters for MAPbI$_3$, as predicted by electronic structure calculations, can be seen in Table 1. The uncertainty in each of the six MAPbI$_3$ input parameters was assumed to be $\pm 20\%$, in the absence of uncertainty estimation accompanying $\emph{ab initio}$ prediction \cite{Irvine21}. This assumed uncertainty results in input parameter values that align with values presented in a range of previous studies, both theoretical predictions and experimental measurements \cite{Frost17, Brivio13, Brivio14, Lee18}. In the hypervolume of the 6-D input space defined by the uncertainty in each input parameter, $300$ points (sets of input parameters) were randomly generated and the polaron mobility calculated at $200$K, $250$K and $300$K. The temperature dependence of the polaron mobility has been characterised by fitting the power-law relationship,
\begin{equation}
    \mu(T) = \mu_0T^{\beta}
\end{equation}
where $\beta$ is the temperature exponent obtained from the gradient of a linear fit performed on log-log axes,
\begin{equation}
    \log(\mu(T)) = \log(\mu_0) + \beta\log(T).
\end{equation}
The temperature exponent $\beta$ was used as the predicted variable $y$ for GPR, as a function of the 6-D input space $\mathbf{x}$. 
As noted in section \ref{subsec:efficient}, Bayesian Optimisation could then be used to minimise the absolute value of the difference between the predicted temperature exponent and an experimentally measured exponent of $-1.5$ (see Figure \ref{fig:mobilitiesvst}). The Python codes used for performing this procedure, with the  supporting data used to run the code, can be found in \cite{Mccallumcodedata23}. Correlations between input parameters have not been considered during optimisation as constraining the optimisation procedure by considering parameter correlations would only decrease the searched parameter space. In this specific application, reducing the searched parameter space may result in a temperature exponent optimum further from the experimental exponent but never closer (as the space considered is inclusive of the correlation constrained space). While any correlations between input parameters will be important in other applications, their inclusion here would not affect any conclusions drawn.

\section{Conclusions}
We have demonstrated that Bayesian Optimisation can efficiently search a region of the simulation input space defined by the uncertainty in each input parameter to minimise the difference between the simulation output and a set of experimental results. This was achieved through explicit evaluation of the probability that the simulation can reproduce the measured experimental results in this region of input space. With this method, 1269 simulations using the code BoltMC were required from 423 points in the 6-D simulation input parameter space (three temperatures at each point in input space). A naive grid search of any simulation input space would require $n^d$ points, for $n$ discrete parameter values along each input dimension $d$, rendering a quantification of input uncertainty computationally intractable for most numerical physics simulations. Here, a coarse grid ($n=10$) would require $2\times 10^6$ simulations. Given the generality of the method, application to other numerical simulations (in semiconductor physics and other fields) is straightforward. For physical models known to exhibit a non-linear functional dependence on model inputs, this method allows for more complete conclusions to be drawn from results as the effect of input parameter uncertainty on the simulation output has been determined.

The minimum temperature exponent of the polaron mobility found during optimisation was $-0.50 \pm 0.04$, with the probability of reproducing an experimental exponent of $-1.5$ found to be $\sim 10^{-24}$. This result suggests that the formation of large polarons in MAPbI$_3$ cannot explain the observed temperature dependence of electron mobility; other mechanisms that regulate charge-carrier dynamics in lead-halide perovskites, such as trapping and recombination, must therefore be investigated.

Additionally, with analysis of the length-scales of the covariance function for GPR, polaron mobility was found to be less sensitive to perturbations in the scattering rates than observed for band electrons. Both band electrons and polarons in MAPbI$_3$ displayed greater mobility sensitivity to perturbations in the scattering rates for polar optical phonons and ionised impurities, with the effect on mobility of a perturbation in the acoustic phonon scattering rate significantly decreased.

\section*{Acknowledgements}
SM and JL thank the UK Engineering and Physical Sciences Research Council (EPSRC) for respectively a summer bursary from Computational Collaboration Project No 5 (CCP5) and a doctoral training partnership studentship.

\section*{Data Availability}
Data that supports the findings of this study is available within the article. The article has no supplementary material. The code with supporting data for this article is publicly available at https://github.com/sammccallum/BoltMC-Bayes-Opt 
where the datasets and code files are described in README.md.
\printbibliography
\end{document}